\documentclass[twocolumn,floats,aps]{revtex4}
\usepackage{mathptmx,amsmath,amsfonts,bm,mathrsfs,psfrag,latexsym,graphics,booktabs}

\begin{document}
\title{Sixty Years of Quantized Circulation}
\author{R.J. Zieve}
\affiliation{Department of Physics and Astronomy, University of California at Da
vis}

\begin{abstract}
Vinen's vibrating wire experiment detected quantized vortices in superfluid
${}^4$He, with the anticipated circulation quantum $h/m$. In addition to this main result, Vinen used his data to propose other properties and behaviors of vortices, which are revisited here. Subsequent work confirmed that non-quantized values occur when vortices cover only part of the wire's length, and that the size of the covered section can change easily. Any motion of the detached portion of the vortex induces changes in the circulation around the wire, which provides a means of tracking the free vortex. Particularly distinctive signatures correspond to a circular motion of the vortex through the cell and to Kelvin waves along the free vortex. Another issue, the lack of stability of multi-quantum states, can also be explained through simple arguments, in which the possibility of a partially detached vortex again plays a key role. Vibrating wire measurements descended from Vinen's continue to probe superfluid flows.
\end{abstract}
\maketitle

\section{Introduction}

More than sixty years ago, Joe Vinen performed a clever experiment that gave
evidence for circulation quantization in superfluid helium \cite{Vinen61}. Previous work, including second sound measurements in rotating superfluid helium by Vinen and Henry Hall \cite{Hall}, was consistent with quantized vortices and difficult to explain otherwise. Vinen's subsequent experiment was far more direct and explicitly measured the circulation quantum $h/m$ predicted by Onsager \cite{Onsager}, where $h$ is Planck's constant and $m$ is the mass of a helium atom. In fact this was the first demonstration of quantization on a macroscopic scale, followed soon after by observations of quantized flux in superconductors \cite{Deaver, Doll}. This article reviews the experiment, discusses subsequent use of the technique, particularly in my own laboratory, and revisits some of Vinen's observations in light of later work.

\section{Vinen's Experiment}

\begin{figure}
\centering
\scalebox{.8}{\includegraphics{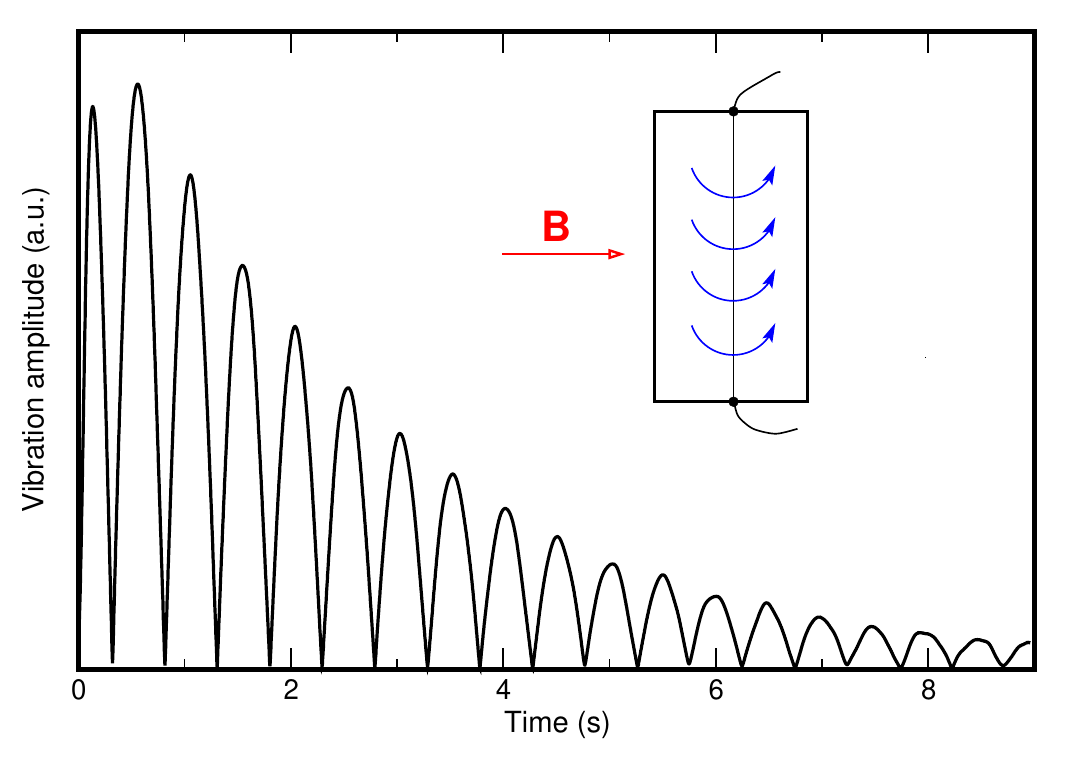}}
\caption{Ringdown of wire immediately after a pulse, near 300 mK. The signal shown is the output of a lock-in amplifier. The wire has a single quantum of circulation around it. Inset: schematic of the wire passing through the cell, with circulation around the wire and a perpendicular magnetic field.}
\label{ringdown}
\end{figure}

Vinen's measurement uses a straight, taut wire immersed in superfluid helium, with a constant magnetic field applied perpendicular to the wire. The inset of Figure \ref{ringdown} illustrates this geometry. A current pulse through the
wire excites vibration. The wire's subsequent motion in the
magnetic field induces an emf across the wire, which is the experimental
signal.

Vinen assumed that upon rotating the fluid, a vortex of one or more quanta would
settle on the wire. If so, then by a standard calculation for the motion of a
straight vortex through a fluid \cite{Lamb}, the wire would feel a Magnus force
perpendicular to its motion. The Magnus force couples motion in the $x$- and
$y$-directions, leading to clockwise and counterclockwise normal modes. These modes have slightly
different frequencies, a natural consequence of symmetry-breaking by the circular flow of the vortex. The beat frequency $\Delta\omega_\kappa$ between the two modes indicates the magnitude of the circulation $\kappa$ around the wire through $\Delta\omega_\kappa=\rho_s\kappa/\mu$. Here $\rho_s$ is the superfluid density and $\mu$ is the mass per length of the wire plus a small correction for the mass of fluid displaced as the wire moves. For the wires Vinen used, a single quantum of circulation would give $\Delta\omega_\kappa=2.8$ rad/s.

Implementing this idea required a great deal of ingenuity. The experiment was by necessity carried out at the relatively high temperature of 1.3 K; many present-day cooling techniques were developed during the following decade. At this temperature the vibrating wire had significant damping from its interaction with the normal component of the helium. As shown in Figure 2 of \cite{Vinen61}, the useful ring time was about 1 second, and only the first minimum of the envelope could be seen. For comparison, Figure \ref{ringdown} shows a ringdown from our much later measurements at 300 mK. Here the emf from the wire has been passed through a lock-in amplifier tuned to the average of the normal mode frequencies, and many beats appear during the much longer measurement time.

Detecting only the first minimum was problematic for Vinen because of real-world imperfection: even in the absence of circulation, the lowest normal modes of a wire are not exactly degenerate, and their splitting is often comparable to the effect of the circulation. It turns out that the two effects add in quadrature:
$$\Delta \omega^2 = \Delta \omega_0^2 + \Delta\omega_\kappa^2$$
where $\Delta\omega$ is the observed beat frequency, $\Delta\omega_0$ is the beat frequency inherent to the wire in the absence of any circulation, and
$\Delta\omega_\kappa=\rho_s\kappa/\mu$ is the beat frequency that circulation $\kappa$ would induce on a perfect wire. Mathematically, it remains simple enough to extract
the circulation, as long as $\Delta \omega_0$ is known and is not so large that it swamps the contribution of $\Delta\omega_\kappa$. Vinen resolved this issue by fastening one end of the vibrating wire to a tube which could be
rotated. The ensuing twist of the wire changed the zero-circulation splitting. In principle the highest sensitivity to circulation comes at $\Delta\omega_0=0$, but there is no way to know that the setup is tuned to this value. With the high damping of the wire, values of $\Delta\omega_0$ with no beats in the observable window could nonetheless distort the calculated circulation. 

\begin{figure}
\centering
\scalebox{.4}{\includegraphics{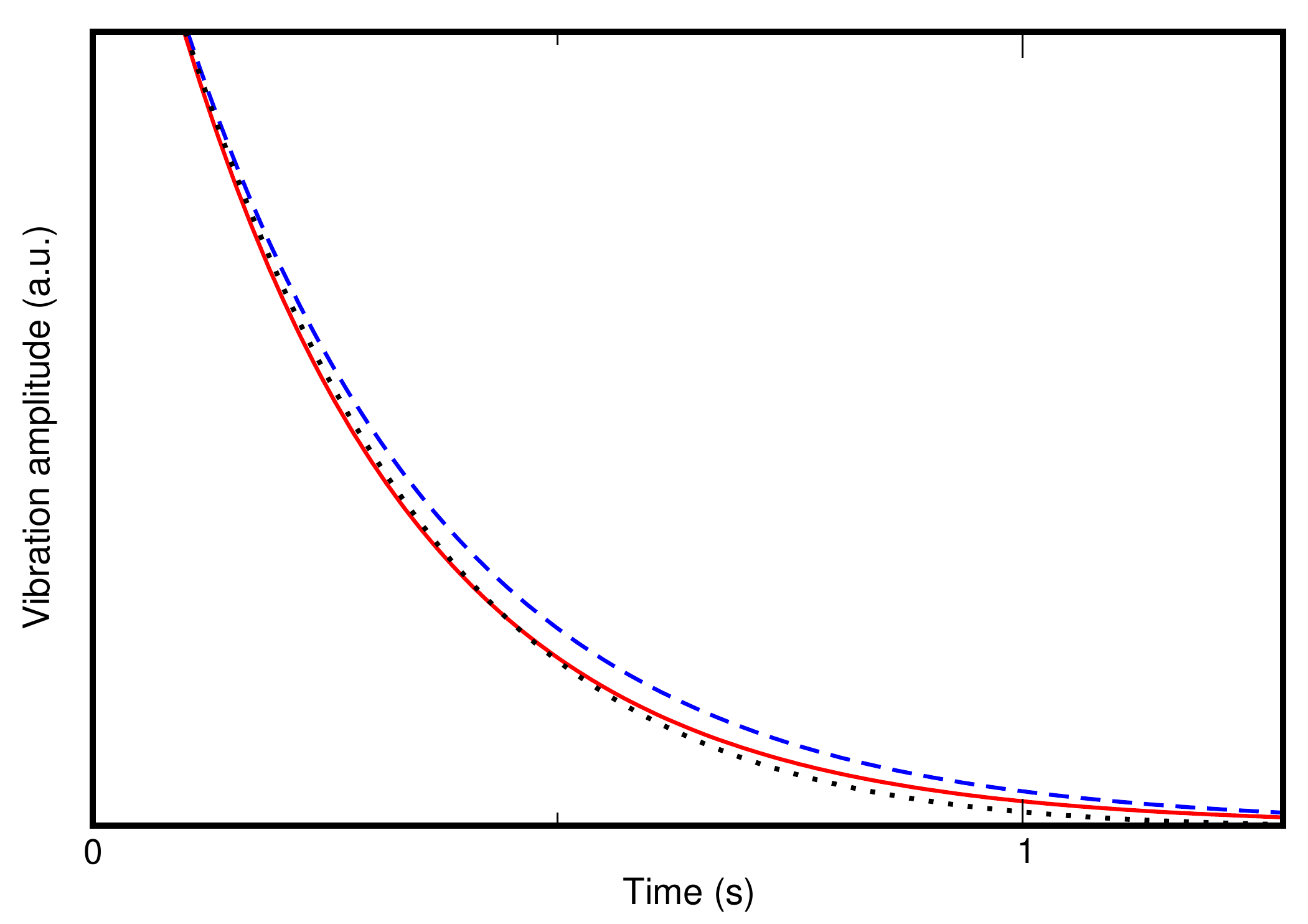}}
\caption{Illustration of the amplitude of a wire's vibration. The damping and the horizontal window of 1.28 s are chosen to match those of Figure 2 in \cite{Vinen61}. Dotted: decay with first minimum at 1.35 seconds, slightly outside the range of the graph. Solid: pure exponential fit to the dotted curve. Dashed: pure exponential, with same time constant as dotted curve.}
\label{zerosplit}
\end{figure}

Figure \ref{zerosplit} illustrates this problem. The horizontal range of 1.28 s is that of Figure 2 from \cite{Vinen61}. The dotted curve is a damped exponential with beats, but with $\Delta\omega_0=2.3$ rad/s so that the first minimum is at 1.35 seconds. The time constant, 0.29 seconds, is also similar to that in \cite{Vinen61}. The dashed curve is a pure exponential with the same time constant but no beats. Finally, the solid curve fits the dotted curve with a single exponential, resulting in a time constant of 0.26 seconds, about 10\% smaller. The imperfection of the fit is apparent in this noise-free example, but would be hidden by typical measurement noise. Furthermore, Vinen had no option of fitting the envelope; his data were photographs of an oscilloscope screen. Thus a wire with beats beyond the measurement window would mimic a perfect wire with slightly higher damping. However, if Vinen had used a wire with $\Delta\omega_0=2.3$ rad/s, then $\kappa$ extracted from the one-quantum level without adjusting for the zero-circulation splitting would be more than 25\% too large, clearly not an acceptable deviation. The only solution was to increase $\Delta\omega_0$ to a measurable level. Vinen opted for a zero-circulation splitting $\Delta\omega_0=3.9$ rad/s, about 40\% higher than the contribution from one quantum of circulation. This gave a first minimum at 0.8 s, near the limit of what could be distinguished reliably.

Another major difficulty lay in deciphering the results. On measuring a quantized phenomenon, the results would ideally show a clear staircase-like structure. That did not happen. The circulation rarely reached the expected two-quantum level. Worse, the circulation often lay in between two of the calculated quantum levels. Vinen interpreted such values as the
result of non-uniform circulation along the wire: a vortex could cover only part of the wire, then continue as a free vortex in the superfluid. He reasoned that such a vortex would be less stable than a vortex covering the entire wire. Hence he monitored the behavior of the measured circulation after a violent excitation of the wire. As anticipated, non-integer circulation nearly always changed significantly, while circulation near the expected one-quantum value tended to remain unperturbed. From this he constructed a histogram of ``stable" circulation, with sharp peaks at 0 and $h/m$ (hereafter referred to as $N=1$). The $2h/m$ level ($N=2$) was rarely seen and was never stable.

Subsequent experiments at lower temperatures, with somewhat altered parameters for the wire, cell, and rotation protocol, found the clean steps Vinen had envisioned \cite{KSZ}. A similar setup was also used to observe quantized circulation in superfluid ${}^3$He, where the fermionic atoms must form Cooper pairs before condensing into a superfluid. Pairing leads to an extra factor of two in the circulation quantum, and the vibrating wire measurement indeed gives a value of $h/2m_3$, with $m_3$ the mass of a ${}^3$He atom \cite{3HeQC}.

As the first measurement with a vibrating wire in superfluid helium, Vinen's circulation work was the forerunner of many subsequent experiments. Most focus on the damping of the wire, which measures viscosity and hence temperature. The first such setup \cite{Tough} used a straight wire much like Vinen's and references Vinen's paper. For a viscometer the main advantage of a straight wire is that the simple geometry allows straightforward calculations of its interaction with the fluid. The near-degeneracy of the normal modes is irrelevant, or even a needless complication, so viscometer designs quickly shifted to hooped wires. These have been used extensively to explore topics as varied as dark matter, superfluid turbulence, ${}^3$He-${}^4$He mixtures, and Majorana fermions \cite{Bradley95, Yano, Pentti, Bunkov}.  After decades of utility, such wires are gradually being supplanted by smaller probes with higher resonant frequencies, such as miniature tuning forks and nanoelectromechanical systems \cite{Bradley17}.

\section{Partly Attached Vortices and Precession}

\begin{figure}
\centering
\scalebox{.68}{\includegraphics{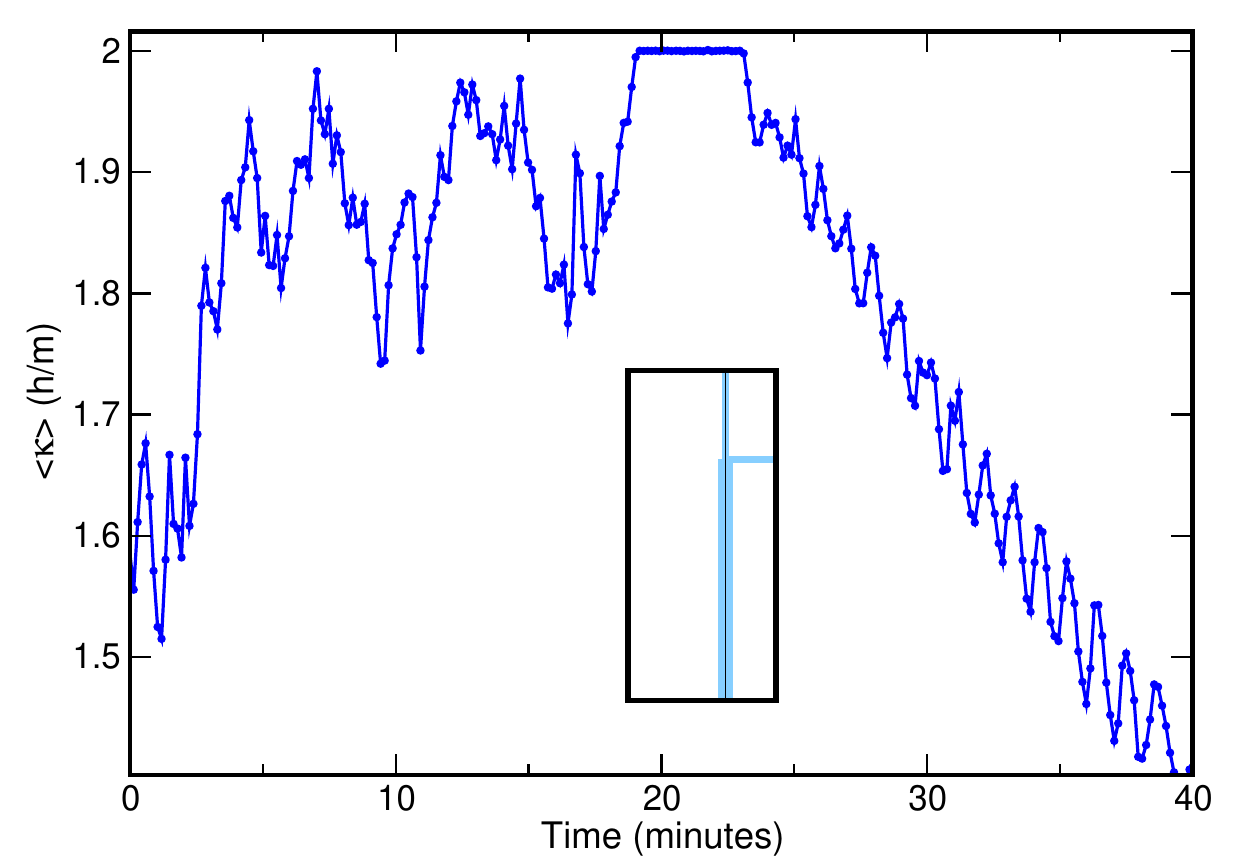}}
\caption{Effective circulation as a function of time at 400 mK, shortly after stopping rotation. Circulation $N=2$ settles on the wire at about 19 minutes. A few minutes later one quantum spontaneously dislodges, followed by precession. Inset: schematic of off-center wire with partially attached vortex, corresponding to the oscillations at right in the main figure. The lower portion of the wire has two quanta of circulation, which split into one along the upper part of the wire and one free vortex terminating on the container wall. As the free vortex moves from the position shown, it lengthens, drawing energy from the trapped circulation and changing the location of the split.}
\label{2to1}
\end{figure}

The most direct descendants of Vinen's experiment \cite{KSZ, helicopter, bump} continued to use a straight vibrating wire to explore vortex behavior in
superfluids. In the process, many of Vinen's suggested interpretations of his data have been confirmed. Notably, he conjectured that intermediate circulation levels indicate vortices which extend only partway along the wire, as illustrated schematically in Figure \ref{2to1}, and that the portion of such vortices not attached to the wire can move through the superfluid. These ideas received striking verification, first in superfluid ${}^3$He \cite{helicopter} and subsequently in ${}^4$He \cite{zievejltp, hough01}. At intermediate circulation the beat frequency often exhibits a regular, very slow oscillation, typically with a period of several minutes, as seen in Figure \ref{2to1} from 25 to 40 minutes. This oscillation never appears at the integer quantum levels. The first piece of the explanation is that free vortices in a fluid tend to move with the local velocity field. In the case of circulation trapped along only part of the vibrating wire, the trapped circulation creates a roughly tangential flow field. The free portion of the vortex moves under the influence of this velocity field, sweeping out a path around the wire. The period of this motion, which can easily be calculated, matches the observed signal and scales correctly in key ways. Increasing the radius of the container decreases the average superfluid velocity, and indeed the precession period increases accordingly. The precession period also depends on whether the circulation is between $N=2$ and $N=1$, as in Figure \ref{2to1}, or between $N=1$ and zero. The period is three times faster in the former case, since the additional circulation increases the fluid velocity. Overall the distinctive signal leaves little doubt that it arises from precession of a partially attached vortex.

There remains the question of why precession of the free vortex induces oscillations in the circulation measurement. The answer lies in the wire's being slightly displaced from the center of the cylindrical cell. With an off-center wire, the free vortex extending from the wire to the cell wall must change length as it precesses, returning to its original length after one cycle around the cell. An oscillation in the kinetic energy of the fluid flow accompanies the length change. This is possible because the portion of the vortex attached to the wire serves as an energy reservoir: as the free vortex lengthens, the attached circulation shrinks, and vice versa. The beat frequency of the wire depends on an effective circulation along its entire length, which increases as a vortex covers a longer segment of the wire. Thus precession of the free vortex translates into an oscillatory signal in the circulation measurements. The attachment point typically moves about $\pm 1$ mm along the wire. This does not mean that the wire is 1 mm from the axis of the cell though. The free vortex, with its tiny core size, has several times greater energy per length than the trapped vortex, so small changes in the free vortex length are amplified to larger length effects on the trapped circulation. Even wire displacements of a few tens of microns -- all but inevitable when stringing a thin wire in place -- give rise to observable oscillations. 

Although partially attached vortices convincingly explain Vinen's observations of intermediate circulation levels, his own measurements did not detect the oscillations described above. There are several possible reasons. First, the expected period given the dimensions of Vinen's cell is about 6 minutes. For each data point he excited the wire, captured the resulting decay on an oscilloscope, and photographed the screen. His published data suggest one measurement about every two minutes, which would not make a 6-minute oscillation very recognizable. Second, his experiment may not have had enough precision, since he could detect only the first minimum in the decay envelope. The subsequent experiments at lower temperatures fit multiple periods to extract the beat frequency. Third, there may have been no precession to observe in Vinen's setup. A vortex terminating at the wall of a container can ``pin" strongly enough that precession ceases. The idea is that the vortex is attracted to bumps on the surface, as expected from hydrodynamic considerations. My group has verified that pinning is more common with rougher cell walls \cite{pinning}, and that a deliberate large bump will reliably pin vortices that approach it \cite{bump}. Furthermore, pinning becomes far more likely with increasing temperature. We repeatedly see vortices that precess below 500 mK, become pinned when the helium is heated above 1 K, and precess freely again once the low temperature is restored. Since Vinen's measurements were at 1.3 K, there may well have been no precession.

\section{Stability of Circulation Levels}

Among the quantized circulation levels, $N=1$ is far more stable than higher levels. Vinen found no indication of other stable levels. Zimmermann's group did see higher circulation in similar experiments, but with larger-diameter wires \cite{WZ} or under rotation \cite{KSZ}. For a stationary cryostat, any non-zero circulation around the wire is only metastable, but my group has observed $N=1$ to remain for more than 60 hours. In fact $N=1$ almost never dislodges without a significant mechanical or thermal perturbation. By contrast, Figure \ref{2to1} shows one of our rare observations of $N=2$. The data begin shortly after the cryostat has been rotated and then brought to a halt. The circulation around the wire fluctuates at first, then settles at $N=2$ for almost five minutes. One end of a vortex then spontaneously leaves the wire, and the familiar precession ensues. The steadiness of the circulation at $N=2$ contrasts sharply with both the fluctuating and precessing regimes, indicating that the double circulation was fully around the wire during that time. 

\begin{figure}[h]
\centering
\hspace*{.2in}
\scalebox{.5}{\includegraphics{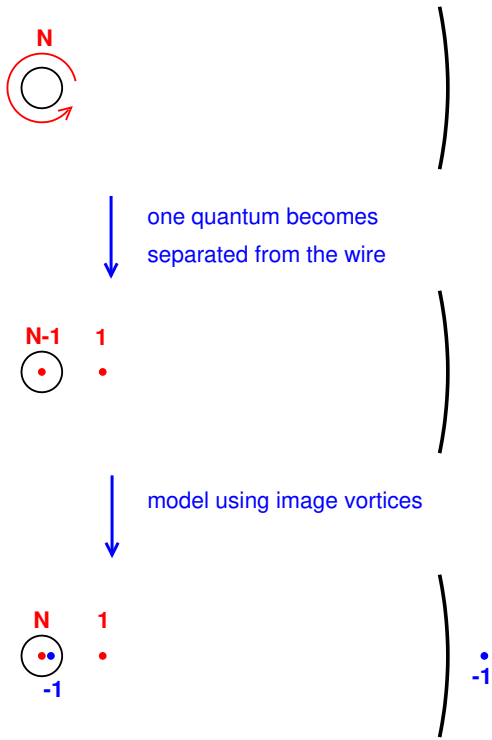}}
\caption{Top view of wire and right part of container wall. Upper: $N$ quanta of circulation on wire.
Middle: possible route for vortex to leave, with $N-1$ quanta on wire and a one-quantum vortex in the fluid. Lower: as described in the main text, we model the resulting flow using image vortices. The objects and locations shown are not to scale.}
\label{images}
\end{figure}

A {\em qualitative} difference between the energetics of $N=1$ and that of higher levels explains the strongly enhanced stability of $N=1$.
We consider $N$ quanta of circulation around a wire of radius $R$ within a cylinder of radius $D$. Following Griffiths \cite{Griffiths}, we explore the energy for one quantum to leave the wire and become a free vortex. For the initial calculation, we assume an effectively two-dimensional geometry, where the free vortex is parallel to the wire at a distance $r$ from the wire's center. As indicated in Figure \ref{images}, we calculate the resulting velocity at the free vortex using image vortices: a -1 vortex (opposite direction to the free vortex) at $D^2/r$, another -1 vortex at $R^2/r$, and a +1 vortex at the center of the wire. The first of these accounts for the distortion of the free vortex velocity field by the outer wall of the cell. The other two account for distortion by the wire, with the central image vortex needed to give the correct behavior far from the wire. Since the container is much larger than the wire, we omit any images of the image vortices and also ignore any displacement of the wire from the center of the container. The superfluid velocity at the free vortex is tangential, with magnitude
\begin{equation}
v_s = \frac{\kappa}{D^2/r -r} - \frac{\kappa}{r-R^2/r} + \frac{N\kappa}{r}.
\label{eq:velocity}
\end{equation}

The velocity at the free vortex leads to a perpendicular Magnus force per unit length $\mathbf{f}=2\pi\rho_s\mathbf{v}_s\times\boldsymbol{\kappa}$, where $\boldsymbol{\kappa}$ has magnitude $\kappa$ and is directed along the vortex core. Integrating this force gives the relative energy of different vortex locations \cite{Griffiths}. The second term in Equation \ref{eq:velocity} contributes a Magnus force directed towards the wire, while the other terms give forces away from the wire. As $r\rightarrow R$ the second term always dominates, creating an energy barrier for a vortex to leave the wire. Physically, the distortion of the velocity field by the wire means that a vortex sufficiently near the wire is always attracted to it.

At larger $r$ the situation depends on $N$. For $N>1$, the third term, the usual repulsion between parallel vortices, takes over. The first term is much smaller and can be ignored, as long as $D$ is much larger than all the other distances. The crossover location, where the velocity and hence the force vanishes, is at $r_b=\sqrt{\frac{N}{N-1}}\mbox{ }R$, which validates dropping the first term. The energy barrier to reach this location, for a vortex of unit length, is
\begin{equation}
\begin{split}
\Delta E=2\pi\rho_s\kappa\int_{R+a_0}^{\sqrt{\frac{N}{N-1}}R} \left(\frac{\kappa}{r-R^2/r}-\frac{N\kappa}{r} \right) dr  \\
\approx \pi\rho_s\kappa^2\ln\frac{(N-1)^{N-1}R}{2N^Na_0},\nonumber
\end{split}
\end{equation}
where $a_0$ is the free vortex core radius and we assume $a_0<<R<<D$. For $r>r_b$, the force on the free vortex is outward and the energy decreases with $r$. 

\begin{table}
\begin{center}
\begin{tabular}{ccc}
\hline
Circulation, $N$\hspace*{.2in} & $\Delta E$ ($10^{-6}$ erg/cm)\hspace*{.2in} & $r_b-R$ ($\mu$m) \\
\hline
1 & 4.9 &  200 \\
2 & 4.3 &  10 \\
3 & 4.1 &  5.6 \\
4 & 3.9 &  3.9 \\
\hline
\end{tabular}
\caption{Energy barrier and location of energy peak if one quantum of circulation out of $N$ leaves a wire of radius 25 $\mu$m in a container of radius 2 mm.}
\label{t:nums}
\end{center}
\end{table}

For $N=1$, the second term in Equation \ref{eq:velocity} always exceeds the third term, with the difference approaching zero for large $r$. This cancellation means that the first term must be retained. To lowest order, the velocity vanishes at $r_b=\sqrt{RD}$, and the energy barrier for a vortex to reach this spot from the wire is $\pi\rho_s\kappa^2\ln(R/2a)$. The picture here is that in an unbounded fluid, a free vortex is {\em always} attracted to a bare wire. Once inside a finite container, the attraction of the free vortex to the outer wall balances the influence of the wire.

\begin{figure}
\centering
\scalebox{.5}{\includegraphics{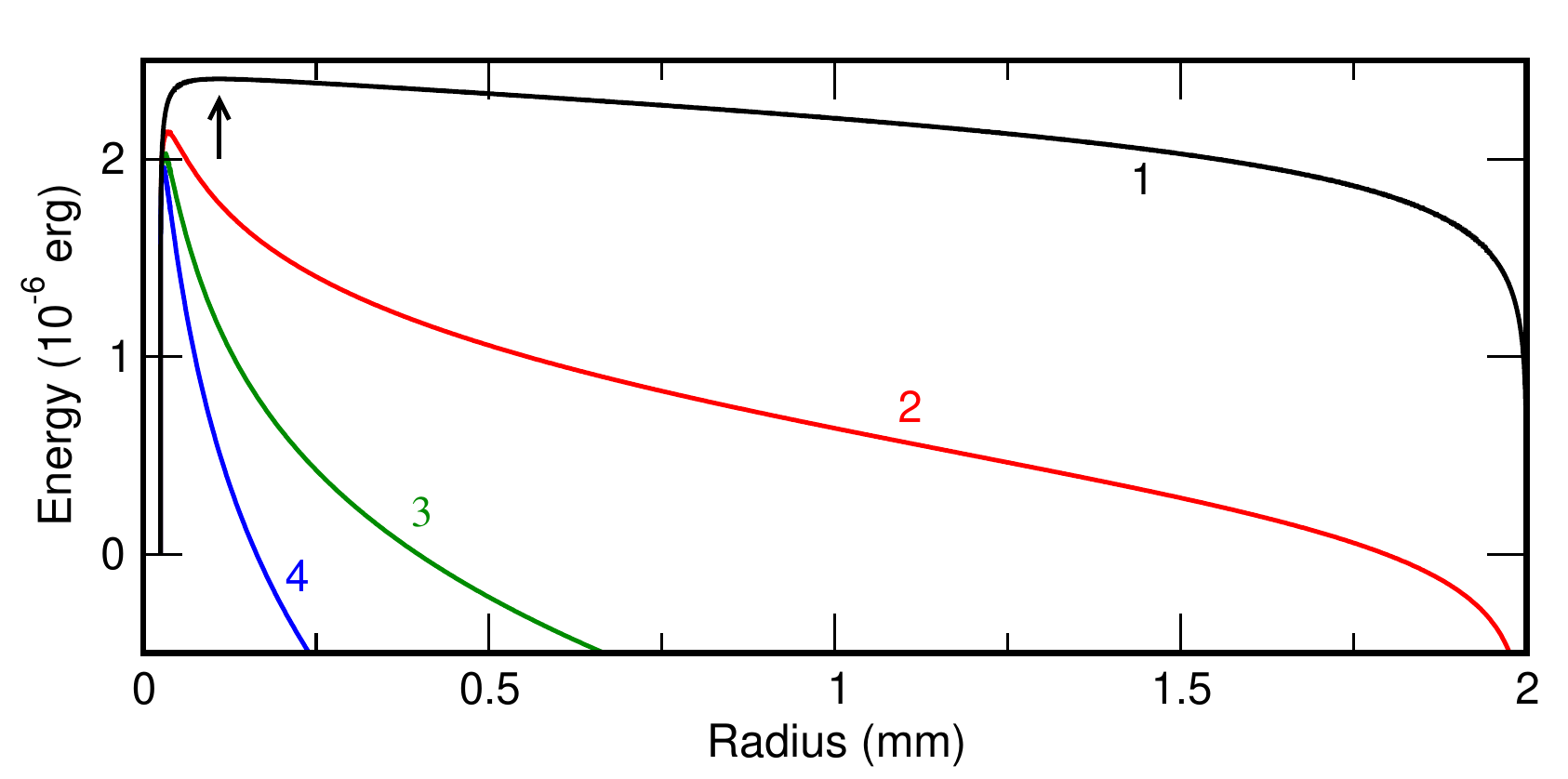}}
\caption{Energy for a single quantum of circulation leaving a wire that initially has $N$ quanta, for $N$ equal 1 through 4. At the zero of energy all the circulation is around the wire. The wire radius is 25 $\mu$m, and it is centered in a cylinder of radius 2 mm. The arrow indicates the maximum energy for the $N=1$ curve.}
\label{energy}
\end{figure}

In Vinen's experiment, $D=2$ mm and $R=25$ $\mu$m. Table \ref{t:nums} gives the corresponding energy barriers and the distances from the wire to $r_b$. Considering only the energies, one might expect comparable stability for these different circulation states. However, the distance from the outer edge of the wire to $r_b$ is twenty times larger for $N=1$ than for $N=2$. In fact, as shown in Figure \ref{energy}, the energy as a function of free vortex position has a far broader peak for $N=1$ than for higher circulation. Since the energy maximum for a vortex leaving a wire with $N>1$ lies very close to the wire, a vortex that moves a short distance away through a fluctuation would then be pulled further away, until it reaches the container wall. With $N=1$ on the wire, the vortex would have to reach a much larger distance to overcome the forces pulling it back towards the wire. 

In reality, a vortex separating from the wire will not maintain the vertical symmetry assumed thus far. In this case the exact velocity field can no longer be obtained from image vortices, but the fundamental difference between the one-quantum and higher states remains. The energy barrier is smaller for a short vortex segment, making it more plausible that a short piece can detach. However, in this case the free vortex cannot be strictly vertical; it needs to extend outward from the surface of the wire as well, adding to its total length and hence its energy. This extra energy contribution is particularly significant for the $N=1$ case because of the much longer horizontal distance required to overcome the attraction to the wire. Another implication of the extra energy term is that the vortex should dislodge near one end of the wire, where only one end of the free segment needs to return to the wire. The precession of Figure \ref{2to1} indeed starts very near the $N=2$ level. Our other observed instances of $N=2$ also dislodge spontaneously, with only a small initial drop from the quantum level.

The experimental observations and above discussion suggest that for $N=2$, one quantum of circulation leaves the wire through detachment of a short portion near one end. After the end of the vortex reaches the outer wall of the cell, precession begins with its accompanying energy loss, gradually reducing the circulation to the $N=1$ level. This mechanism appears not to act on the $N=1$ state, perhaps because of the different shape of the energy barrier. For $N=1$, fluid flow or mechanical perturbation can dislodge the vortex from the wire, but the details of these interactions remain unknown.

\section{Kelvin Waves}

\begin{figure}
\centering
\scalebox{.4}{\includegraphics{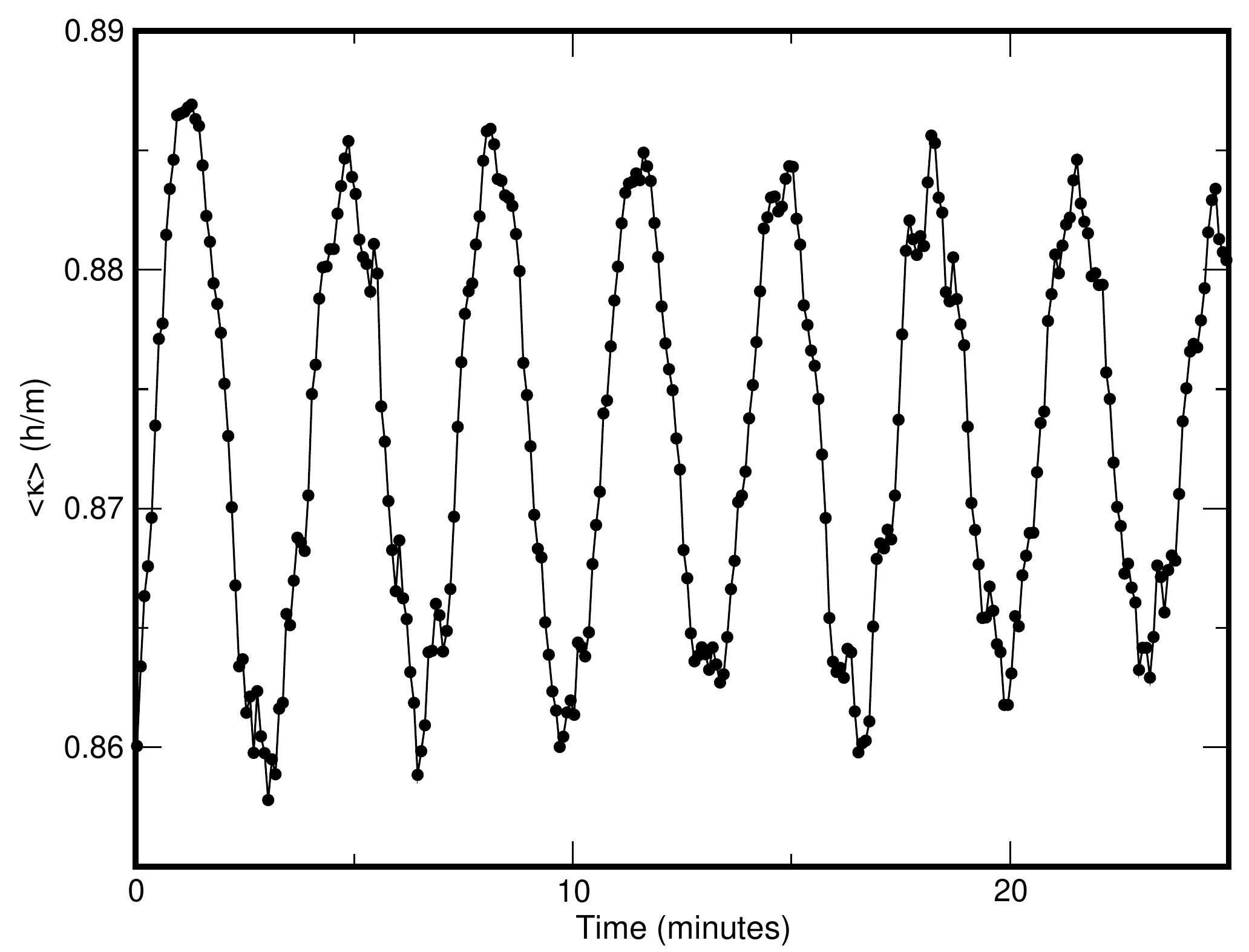}}
\caption{Effective circulation as a function of time, showing Kelvin waves excited along a pinned vortex. The lack of change in the average circulation indicates the pinning.}
\label{kelvinwave}
\end{figure}

When circulation extends along only part of the wire, the vibrating wire technique provides a unique way to probe the behavior of the resulting free vortex segment. The characteristic oscillations from a precessing vortex are one example. Another is detection of Kelvin waves along the free vortex \cite{donev01}, as illustrated in Figure \ref{kelvinwave}. Here one end of the free vortex is pinned at the wall, so the average effective circulation remains constant. The other end of the vortex moves along the wire, which changes the length of trapped circulation and produces the oscillatory signal. The lowest mode meeting these restrictions has wavelength $\lambda$ four times the length of the free vortex.
Using the cell radius of 3.7 mm and calculating frequency from
$$\omega = -\frac{\kappa \pi}{\lambda^2}\ln\frac{a}{\lambda}$$
gives 26 rad/s, not far from the observed 31 rad/s. This cell was set up with the wire deliberately off-center, so the discrepancy probably means that the free vortex extends in a direction where the wire is slightly closer to the cell wall. A length of 3.4 mm would match the measured frequency. The wire itself excites the Kelvin waves as it vibrates, dragging one end of the free vortex along with it.

One question is, why does only the lowest mode appear? The wire's vibration, near 447 Hz, should excite higher waves at least as easily. The next mode, with 3/4 of a wavelength along the free vortex, would have a period near 24 seconds, which could be detected at the sampling rate of Figure \ref{kelvinwave}. Beyond that, the modes would be too fast, but they should appear aliased to a lower frequency. Even if their motion is somewhat irregular, they should contribute an apparent noise. Yet from Figure \ref{kelvinwave}, any other Kelvin waves have significantly smaller amplitudes. For a possible answer we return again to Vinen \cite{Vinen03}. With Tsubota and Mitani, he modeled a helium vortex with continuous excitation of a single Kelvin mode. The numerical results show nonlinear coupling that transfers the energy to other modes. In the eventual steady-state spectrum, the amplitude of a mode is proportional to $\lambda^{1.5}$. Our apparatus may set up exactly such an energy transfer, with the higher modes not detected because of their reduced amplitude. 

Kelvin waves are key to low-temperature superfluid turbulence but also govern other aspects of vortex behavior. For example, Vinen was intrigued with the observation in \cite{donev01} that pinning increases at higher temperatures. The opposite would be expected if vortices work free from pin sites through thermal excitation. This occurs in superconductors, where vortices can ``creep" between pinning sites through thermal activation, with motion increasing with temperature \cite{Kwok}. In our experiment, pinning of the vortex at the cell wall is instead overcome through the Kelvin waves excited by the wire's vibration. Pinning events become more common whenever the Kelvin waves are less pronounced, including at higher temperatures or with less frequent or less intense excitation of the wire.

\section{Current Directions}

We are pursuing further investigations and have observed the second Kelvin mode when using very strong excitations. We are now focusing primarily on the Kelvin wave spectrum and on the nonlinearities that cause energy transfer between the modes. A principal advantage of the vibrating wire apparatus is the ability to set up repeatably a well-defined and non-trivial vortex configuration, notably a vortex stretched between the wire and a specific location on the cell wall. The main deficiency is the restriction to low-frequency signals. Combining the wire with a higher-frequency detector may extend Vinen's vibrating wire technique even further.

\section{Acknowledgments}

The author thanks the Laboratoire de Physique des Solides of the Universit\'{e} Paris-Saclay for hospitality during the writing of this paper.

\end{document}